%% file: main.tex
\pdfoutput=1
\documentclass[10pt,journal,compsoc]{IEEEtran}
\usepackage{dblfloatfix}
\usepackage{balance}
\usepackage{alltt}
\usepackage{pifont}

\usepackage{graphicx}
\usepackage{colortbl}
\usepackage[dvipsnames]{xcolor}
\usepackage{rotating}
\usepackage{makecell}
\usepackage{tabularx}
\usepackage{booktabs}
\usepackage{wrapfig}
\usepackage{tikz}
\usetikzlibrary{angles}
\usepackage{multirow}
\usepackage{hyperref}
\usepackage{framed}
\usepackage[framemethod=tikz]{mdframed}
\usetikzlibrary{shadows}
\usepackage{enumitem}
\usepackage{graphics}
\usepackage[para,online,flushleft]{threeparttable}

\setlist[description]{leftmargin=1cm}

\usepackage{pgfplots}
\usetikzlibrary{patterns}
\usetikzlibrary{arrows}
\usetikzlibrary {positioning}
\usepackage{enumitem}
\setlist[itemize]{leftmargin=*}
\setlist[enumerate]{leftmargin=*}
\usepackage{makecell}
\usepackage[linesnumbered,ruled,vlined]{algorithm2e}
\setlist{nolistsep} 

\SetKwProg{Fn}{Function}{}{}

\setlist[1]{itemsep=0pt}

\newmdenv[
tikzsetting= {fill=gray!10},
linewidth=1pt,
roundcorner=2pt,
shadow=false
]{myshadowbox}

\newcolumntype{P}[1]{>{\centering\arraybackslash}p{#1}}

\hypersetup{
    linkcolor=blue,
    filecolor=magenta,      
    urlcolor=cyan,
}

\newcommand{\bi}{\begin{itemize}[leftmargin=0.4cm]}
\newcommand{\ei}{\end{itemize}}
\newcommand{\be}{\begin{enumerate}[leftmargin=0.4cm]}
\newcommand{\ee}{\end{enumerate}}
\newcommand{\fig}[1]{Figure~\ref{fig:#1}}

\newcommand{\tbl}[1]{Table~\ref{tbl:#1}}

\newcommand{\tion}[1]{\S\ref{tion:#1}}

\usepackage{url}

\tikzstyle{thmbox} = [rectangle, rounded corners, draw=black, fill=gray!10]

\usepackage[tikz]{bclogo}
\newenvironment{RQ}[1]%
{\noindent\begin{minipage}[c]{\linewidth}%
\begin{bclogo}[couleur=gray!25,%
                arrondi=0.1,%
                logo=\bctrombone,%
                ombre=true]{{\normalsize ~#1}}}%
{\end{bclogo}\end{minipage}\vspace{2mm}}

\usepackage{listings}
\usepackage{cellspace}
\ifCLASSOPTIONcompsoc
  \usepackage[nocompress]{cite}
\else
  \usepackage{cite}
\fi

\ifCLASSINFOpdf
\else
\fi

\newcommand{\review}[1]{{\textit{#1}}~\\}

\newcommand{\revised}{\textcolor{black}}

\newcommand{\citeresp}[1]{
{\bf (see } \fcolorbox{black}{black!15}{
 \bf
  \scriptsize R-{#1}}~{\bf{on page \pageref{response:#1})}}
}

\begin{document}

\title{Better Data Labelling with  EMBLEM\\ (and how that Impacts Defect Prediction)}


\author{Huy Tu, Zhe Yu, Tim Menzies~\IEEEmembership{IEEE~Fellow}
\thanks{Authors are from   NC State (Tu, Menzies), and Rochester Institute of Technology (Yu), USA. hqtu@ncsu.edu, aaazhe825@gmail.com, timm@ieee.org}}


\IEEEtitleabstractindextext{%
\begin{abstract}
Standard automatic methods for recognizing
problematic development commits can be  greatly improved  via the
incremental application of
 human+artificial expertise.
 In this approach,  called EMBLEM, an AI tool first explore the software development process to label commits
 that  are most problematic.  Humans then apply their expertise to  check those labels (perhaps resulting in the AI updating the  support vectors within their SVM learner). We recommend this human+AI partnership, for several reasons.
When a new domain is encountered, EMBLEM can learn better ways to label which comments refer to real problems. Also, in studies with 9 open source software projects, labelling via EMBLEM's incremental application of human+AI is at least an order of magnitude cheaper than existing methods ($\approx$ eight times). Further, EMBLEM is  very effective. For the data sets explored here, EMBLEM better labelling methods  significantly improved
$P_{opt}20$ and G-score performance in nearly all the  projects studied here.

\end{abstract}
 
}

\maketitle

\IEEEdisplaynontitleabstractindextext
\IEEEpeerreviewmaketitle

\ifCLASSOPTIONcompsoc
\IEEEraisesectionheading{\section{Introduction}\label{sec:introduction}}
\else
\section{Introduction}
\fi

 \IEEEPARstart{A}{} vital first step in software analytics  is finding the labels 
(or  “ground truth”) for training data. It can be very expensive to acquire these labels via human labor.
For example,
four out of the nine projects studied in this paper have $22,500+$ commits which
required 175 person-hours, include cross-checking, to read
via standard manual methods (and 175 hours $\approx$ nine weeks of work).

Since labelling is so labor-intensive, researchers often reuse
datasets labelled from previous studies. For instance, Lo et al., Yang et al., and Xia et al. certified their methods using data generated by Kamei et al. \cite{xia16ist17, yang16unsupervised, yang2015deep}. While this practice lets  researchers rapidly
test new methods, it does mean that 
bad labels (that were mistakenly
assigned in prior work)
can cascade over an entire research community.

Another standard practice in defect prediction ~\cite{commitguru, Kim08changes,catolino17_jitmobile,nayrolles18_clever,mockus00changeskeys,kamei12_jit,hindle08_largecommits} is to label a commit as ``bug-fixing''
when the commit text contains   keywords  like those of Table~\ref{tab:words}.
Vasilescu et al. \cite{Vasilescu15github,Vasilescu18z} notes that these keywords are used in somewhat ad hoc manner 
(researchers peek at a few results, then tinker  with regular expressions that combine these keywords). 
That ad hoc approach can perform very
badly. The `Keyword' column of Table \ref{tbl:sample} shows what happens
when we performed the standard labelling on one corpus of projects.
Note that the keyword method performed very badly to identify bug-fixing commits.

 Table \ref{tbl:sample} also shows that the same commits were labelled accurately by EMBLEM, the incremental support vector machine (SVM) approach introduced in this paper. 
 EMBLEM works by combining AI tools (to find potentially interesting examples)
 with human expertise (that comments on those examples).
EMBLEM    reduces the costly efforts associated
 with humans labelling a large corpus
 by guiding humans 
 to read the fewest examples
 needed to best train a classifier to capture most of the important (i.e. bug-fixing) commits. 

\begin{table}[!t]

 \caption{This paper argues {\em against} using   keywords like these as a method for labelling a commit as ``buggy'. Keywords from~\cite{hindle08_largecommits}. }\label{tab:words}

\begin{tabular}{l|l}
\rowcolor{gray!30}      \textbf{Category}  &    \textbf{Associated Keywords}    \\\hline
Corrective &  bug, fix, wrong, error, fail, problem, patch \\
Feature Addition & new, add, requirement, initial, create \\
Merge &     merge \\
Perfective & clean, better \\  
Preventive & test, junit, coverage, asset 
\end{tabular}
\vspace{-5pt}
\end{table}
   


\input{sample_commits.tex}
The process of linking commits to their associated changed code
has been well studied~\cite{Sliwerski05changes, costa17szz, Kim08changes, lo19_mislabeled, Wu11_relink, Tian12_linux}. But once that link is created, it must be {\em labelled}
as ``bug fixing'' or otherwise. As shown below, this problem (a)~takes much effort; and (b)~that effort can be significantly reduced via the methods of this paper. To demonstrate that,   we ask the following questions.
\vspace{1pt}

\noindent
{\bf RQ1: How close are EMBLEM and keyword labelling to human labels?}
This question compares different labelling methods (keywords, EMBLEM)
against ground truth labels (assigned by a team of humans).
Here, we show that: 
\begin{RQ}{\normalsize{Compared to humans...}} 
EMBLEM was best at reproducing the ground truth (i.e. the human labels).
\end{RQ}



\noindent
{\bf RQ2: Does keyword labelling lead to better predictors for buggy commits? }
Here, we compared the effectiveness of various data miners (LR, RF, SVM, and FFTs) and then chose the best one to test the effectiveness of the labelling method (keywords or EMBLEM).
Project's data is divided into releases,
classifiers were trained on prior releases 
(and tested on subsequent releases).
Here, we show that:

 \begin{RQ}{\normalsize{Compared to keyword labelling...}} 
 EMBLEM generated better predictors for bugs.
\end{RQ}

\noindent
{\bf RQ3: How much effort is saved by EMBLEM? }
We compare the time required to label   commits from 50 projects via EMBLEM of
manual means. Assuming
 we were paying Mechanical Turk workers to perform that labelling, then
  manual labelling would cost \$320K and 39,000 hours (assuming pairs of workers per commit, and a  50\% cull rate for quality control).
Using EMBLEM, that same task would  cost \$40K and 4,940 hours; i.e. it would be   8 times cheaper 
(for details on how we make this calculation, see Table \ref{tbl:time1} and Table~\ref{tbl:time2}). That is: 
\begin{RQ}{Compared to manual labelling...} 
EMBLEM-based labelling can be an order of magnitude cheaper.
\end{RQ}

In summary, for defect prediction, we recommend EMBLEM. 
As to the novel contributions of this paper:
\be
\item We demonstrate the prime importance of labelling data as a  vital task in the analytics pipeline (this is a contribution since, previously, this issue has received scant attention).
\item To the best of our knowledge, this work is the first to use active learners to generate labels for defect prediction. Instead of using keywords or training the model to predict bug-fixing commits (as done
in standard approaches), our active learning framework (EMBLEM) built models
from human expertise to better sampling which instances should be
labelled next.
\item We show that even when starting
with zero prior knowledge, this incremental
defect label methods lets us build
useful models, after inspecting only a small portion of the commit logs. As shown in \tion{iia}, this method can reduce the cost of labelling commits by an order of magnitude (by approximately eight times).
\item To better support other researchers our scripts and data are on-line at github.com/sillywalk/defect-prediction/.
\ee

The rest of this paper is structured as follows. Background work is discussed in the next section. 
\S3 and \S4 describes our empirical methodology and experimental design. This is followed by the details of the experiment used
to answer our research questions in \S5. 
Further discussion, threats to validity, and possible
future work from this research are explored in \S6, \S7, and \S8. Finally, the conclusions of this work are given in \S9.

\begin{figure*}[!t]

\centering \includegraphics[width=0.95\linewidth]{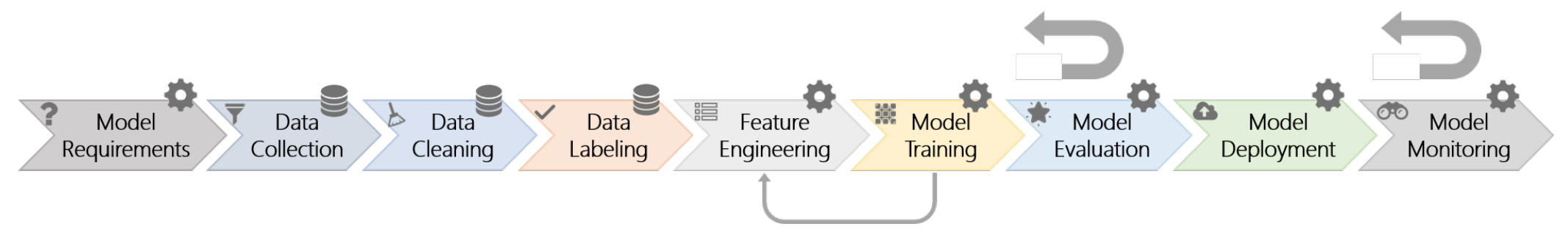}
\caption{Nine stages of the machine learning workflow from a case study at Microsoft by Zimmermann et al. \cite{amershi_ICSE19AI4SE}. Some stages are data-oriented (e.g., data collection, cleaning, and labelling) and others are model-oriented
(e.g., model requirements, features engineering, model training, evaluation, evaluation, deployment and monitoring).}\label{fig:pipeline}
\vspace{5pt}
\end{figure*}   

\section{Motivation and Background}

\subsection{Data Sharing: Benefits and Drawbacks}

Most of this paper discusses problems (and solutions)
of data labelling for defect prediction.
Before that, this section steps back and comments that the methods
discussed here might actually have a broader application area.
Specifically, we think that our methods are an important contribution to the general area of data sharing and reproducibility in software engineering (in particular) and science (in general).

Standard practice in the software analytics literature is for different researchers to try their methods on shared data sets.
For example, in 2010, Jureckzo et al.~\cite{Jureczko:2010} offered tables of data that summarized dozens of open source JAVA projects.
That data is widely used in the literature. A search at Google Scholar on ``Xalan synapse'' (two of the Jureckzo data sets)
shows that these data sets are used in 177 papers and eight textbooks, 126 of which are in the last five years.

Reusing data sets from other researchers has its advantages and disadvantages.
One advantage is {\em repeatability of research results}; i.e. using this shared data, it is now possible and practical
to repeat/repute/prove prior results. For examples of such kind on analysis, see the proceedings of the PROMISE conference
or the ROSE festivals (recognizing and rewarding open science in SE) at FSE'18, FSE'19. ESEM'19 and ICSE'19. 
See also all the lists of 678 papers that reuse data from the Software-artifact Infrastructure Repository at Nebraska University (sir.csc.ncsu.edu/portal/usage.php).

Another advantage is {\em faster research}.
Software analytics data sets contain independent and dependent variables. For example,
in the case of self-admitted technical debt (SATD), the independent variables are the programmer comments and the dependent variable
is the label ``SATD=YES'' or ``SATD=NO''.
Independent variables can often be collected very quickly
(e.g. Github's API permits 5000 queries per hour).
However,
assigning the dependent labels is comparatively a much slower task.
According to Maldonado and Shihab et al.~\cite{maldonado2015detecting}, classifying 33,093 comments as ``SATD $\in$ \{yes,no\}''
from five open source projects took approximately 95 hours by a single person; i.e. 10.3 seconds per comment.
Using that information, we calculated that, relabelling the data used in this paper would require
months of work (see \S2.2 and Table~\ref{tbl:time1} with Table~\ref{tbl:time2}).
When a task takes months to complete,
it is not surprising that research teams tend to reuse old labels rather
than make their own.

That said, the clear disadvantage of reusing old labels is {\em reusing old mistakes}.
Humans often make mistakes when labelling~\cite{hatton2008testing}. Hence, it is prudent to review the labels found in the dataset.
As we will show here: (a)~manually reviewing software artifacts
is a complex process; and (b)~much of that cost can be reduced via a partially-automated procedure, such as the EMBLEM system discussed here. 

Of course, labelling is only one part of the whole process
of building models from data. Figure \ref{fig:pipeline} offers one
description of that whole process and the next section of this paper discusses
where labelling fits into that picture.  

\subsection{The role of labelling in data mining}\label{tion:iia}

One of the goals of industrial analytics is that new conclusions can be quickly obtained from new data just by applying data mining
algorithms.

As shown in Figure \ref{fig:pipeline}, there are at least nine separate
stages that must be completed before that goal be reached~\cite{amershi_ICSE19AI4SE}. Each of these stages offers
unique and separate challenges, each of which deserves extensive attention. They are mapped to our problem as:
\vspace{5pt}
\begin{enumerate}[start=1,label={\bfseries Step \arabic*}]

\item \textbf{Data Collection}: existing organizational data stores are queried for their relevant data.

\item  \textbf{Bug-fixing Labelling} (Data Labelling): categorize a commit as bug-fixing or not based on the textual content of the commit log. 

\item  \textbf{Bug-inducing Identification} (Feature Engineering Part 1): determine the commit that inducing the bugs from the bug-fixing code change what code changes were implicated in a bug report (i.e. SZZ algorithm, which is short for Sliwerski, Zimmermann, and Zeller \cite{Sliwerski05changes}). 

\item  \textbf{Feature Engineering} (Feature Engineering Part 2): describe code change as a set of features and apply any data transformation as necessary (e.g. regularization, sampling). Recent work utilized the deep belief network to generate more quality metrics \cite{yang2015deep}.

\item  \textbf{Model Training}: decide which data miners to use, then apply it to the data.

\item  \textbf{Model Evaluation, Deployment, and Monitoring}: apply different metrics to assess the effectiveness of the data mining process while juxtaposing the result with the hypothesis. Outside of numerical performances (e.g. precision, recall, F-measures), it is also as important to evaluate the explainability of the models (e.g. through visualization) to have insights on when and how models fail to make accurate predictions \cite{amershi_ICSE19AI4SE}.
 
\end{enumerate}

Many of these steps in Figure \ref{fig:pipeline} have been extensively studied in
the literature\cite{ghotra15, menzies07dp,xia16ist17, yang16unsupervised, yang2015deep, Kim08changes, Sliwerski05changes, costa17szz}. However, the labelling work of step 2 of the revised framework mentioned above (or step 4 from Figure \ref{fig:pipeline}) has been received scant attention. Some of the prominent representatives for this area can be traced back to 2010 and 2012 and include:
\be
\item Linkster \cite{linkster_10}: Bird et al. proposed a tool that enables
experts to quickly find and examine relevant changes, and annotate them as desired by integrating multiple queryable, browseable, time-series views of version control history and bug report history.
\item Relink \cite{Wu11_relink}: this approach begins with the labelled bugs then it does the linking analysis to match the changes to the bugs. It generates the labels and validates the veracity of those in the first place. Their methods are not valid and applicable unless the labelling is correct in the first place (which needs EMBLEM).
\item Lo et al. \cite{Tian12_linux} proposed a combination of Learning from Positive and Unlabelled Examples (semi-supervised) and SVM (supervised) to identify Linux bug-fixing patches. 
\ee 
Note that Linkster helps researchers manually label around 500 commits within a working day. Assuming that one working day is 8 hours, it takes approximately 1 minute
to study each commit. Hence, a large scale Linkster-style analysis would be too costly in real-world
scenarios. More importantly, Linkster assumes the existence of mailing list data (that our study does not have access to).

As to Relink and the methods of Lo et al., these
approaches assume the existence of some pre-labelled examples.
For example, suppose we wish to certify the effectiveness of the Lo et al. method.
Such a certification would require a library of ground-truth examples that were curated and certified efficiently by some smart processes. 
This paper proposes and discusses that cost-effective and smart method for generating such a library of ground-truth examples, EMBLEM as the human+AI framework.


More generally, we argue that it is both
{\em necessary} and {\em pragmatically useful}
 to study labelling generally and commit labelling for defect prediction specifically.

{\em (1)~Necessary: } The second and third column of
Table \ref{tbl:gittracking} show that the relationship between code-fixing commits
and issues reports.  Note that there is not a straight-forward linking from one to another.
In our experience,  it is an inherently
complex and error-prone task to generate those links.
For example,  the last two columns of
Table \ref{tbl:gittracking} shows our analysis of (a)~the links from code
commits to issues raised by the physicists
and (b)~the links from those issues back to related code commits
(these two columns were generated by text mining the commit comments and the text of the issue reports).
If the commit comments (or the text of the issue reports) covered the matters raised in columns two and three,
 then the percentages in columns should be large numbers.
 The key thing to observe in that table is: 
\begin{itemize}
\item The percentage of commits being linked back successfully to the issues and the percentage of issues can be inferred from commits (shown in columns four and five in the table) is shown to be \textbf{very low} indeed (down to 1\% for some cases with median of 3\%-9\%).
\end{itemize}
 When data looks like Table \ref{tbl:gittracking}, there are no effective automated approaches and  researchers are forced
to check   {\em all} commits and relabel all those that
actual refer to bug fixes. It would be unnecessary to check all commits with the approach proposed in this paper, i.e. EMBLEM.


{\em (2)~Pragmatically Useful: } 
Checking and relabelling is a time-consuming and expensive task. While most commit messages are short\footnote{The sample in \tbl{sample} reflects the median size of these messages.},  the nine projects studied in this paper have $45,000+$ (before pre-processed) commits. Based on experience with running data-labelling sessions (using teams of graduate students), labelling the  $22,500+$ commit messages from these projects required 175 person-hours (or 19.5 hours on average for a project; median reading
times see in $5000+$ commits). These nine projects are just a sample of the 59 computational science projects that
we have currently found on Github (and we suspect that an order of magnitude more such projects may exist).
Assuming these 59 projects have the same commit frequency as the nine, then labelling all 59 projects requires 29 weeks of work. Further, if a second human is used to check the labels (which is a standard practice in manual SE research papers), this estimate grows to 58 weeks (1.15 years). And we are not finished yet. If any other research team wants to check our results (which is always good research practice) then yet another 1.15 years may be required (per secondary group), just to independently check those labelling.

\begin{table}[!t]
\caption{How often can the bug-fixing commits be linked back to the specific issues through issue tracker (from Github). See \tbl{data} for notes on the projects shown in column one.}
\scriptsize
\begin{center}
\vspace{-5pt}
\begin{tabular}{r@{~}|r@{~}|r@{~}|r@{~}|r@{~}}
\rowcolor{gray!30}  & No. of &        & Commits & Issues Linked      \\
\rowcolor{gray!30} Project & Bug Fixing & No. of & Linked to  & Through     \\
\rowcolor{gray!30}  & Commits  & Issues & Issues \% & Commits \%\\\hline
RMG-PY & 609 & 1698 & 19 (3\%) & 16 (1\%)    \\ 
LIBMESH & 1720 & 2263 & 56 (5\%) & 30 (1\%) \\
LAMMPS & 664 & 1702 & 84 (9\%) & 29 (3\%) \\ 
AMBER & 815 & 763 & 11 (5\%) & 20 (3\%)  \\ 
HOOMD & 1124 & 470 & 41 (3\%) & 13 (3\%)  \\ 
PCMSOLVER & 198 & 178 & 18 (9\%) & 13 (7\%)  \\ 
MDANALYSIS & 707 & 2360 & 209 (29\%) & 193 (8\%)   \\  
XENON & 507 & 657 & 128 (25\%) & 135 (20\%)   \\ 
ABINIT & 730 & 25 & 64 (9\%) & 15 (60\%)   \\  
\end{tabular}
\label{tbl:gittracking}
\end{center} 
\end{table}



What would it cost to complete all that labelling? The following estimates assume
(a)~the use of crowdsourcing (via Mechanical Turk);
(b)~our crowdworkers are being paid at least minimum age (\$8.25); 
(c)~we assign two readers per issue report; and 
(d)~a 50\% ``cull rate'' of crowd workers (where quality control questions are used to identify and prune ineffective crowdworkers\footnote{Such a 50\% ``cull rate'' is common practice in crowdsourcincg~\cite{chen19}};
and (e)~and our university takes a 50\% overhead tax on grants. Under those assumptions, labelling 500 projects of Github issues would require 39,000 hours and \$320K of grant money (with nothing left over for graduate student wages or other equipment).

However, when using EMBLEM, we would only need to read up to 22\% on the median of the commits to find 95\% of the bug-fixing commits (documented in the last column of \tbl{data}c).
That is, the same task would only consume 4,940 hours and \$40K of the money grant (which is an order of magnitude improvement by approximately eight times). For details on how we make this calculation, see Table \ref{tbl:time1} and Table~\ref{tbl:time2}, later in this paper.

\subsection{Why Study Defect Prediction?}

The case study of this paper relates to defect prediction. This section motivates
the values of studying this part of software engineering.

 Software quality assurance budgets are finite while assessment effectiveness increases exponentially with assessment effort~\cite{fu2016tuning}. Therefore, standard practice is to apply slower methods on code sections that seem most critical or bug-prone.
Software bugs are not evenly distributed across the project~\cite{hamill2009common,koru2009investigation, ostrand2004bugs,misirli2011ai}. Hence, 
a useful way to perform software testing is to allocate most assessment budgets to
the more defect-prone parts in software projects.  Data mining algorithms can input
features extracted from source code and output predictors for where defects are likely to occur.
Which such predictors are never 100\% correct,
they can suggest where to focus on more expensive methods.

There is much commercial interest in defect prediction. In a survey of 395 practitioners from 33 countries and five continents,
Wan et al.~\cite{wan18} found that over 90\% of the respondents were willing to adopt defect prediction techniques. 

 Results from commercial projects have shown the benefits of defect prediction.
Misirli et al.~\cite{misirli2011ai} built a defect prediction model for a telecommunications company. Their models predicted 87\% of code defects and decreased inspection efforts by 72\% (while reducing post-release defects by 44\%). Kim et al.~\cite{kim2015remi} applied the defect prediction model, REMI, to the API development process at Samsung Electronics.
Their models could
predict the bug-prone APIs with reasonable accuracy~(0.68 F1 scores) and reduce the resources required for executing test cases.

Software defect predictors not only save labor compared with traditional manual methods, but they are also competitive with certain automatic methods. 
Rahman et al. ~\cite{rahman2014comparing} compared (a) static code analysis tools FindBugs, Jlint, and PMD with (b) defect predictors (which they called ``statistical defect prediction'') built using logistic regression.
No significant differences in cost-effectiveness were observed.

Given this equivalence, it is significant to
note that defect prediction can be quickly adapted to new languages by building lightweight parses to extract code metrics. The same is not true for static code analyzers - these need extensive modification before they can be used in new languages.
Because of this ease of use, and its applicability to many programming languages, defect prediction has been  extended many ways including:
\be
\item Application of defect prediction methods to locating code with security vulnerabilities~\cite{Shin2013}.
  \item Predict the location of defects so that appropriate resources may be allocated (e.g.~\cite{bird09reliabity})
  \item Understand the factors that lead to a greater likelihood of defects such as defect prone software components using code metrics (e.g., ratio comment to code, cyclomatic complexity) \cite{menzies10dp, menzies07dp, ambros10extensive} or process metrics (e.g., number of changes, recent activity) \cite{nagappan05codechurn,elbaum00codechurn, moser08changemetrics, hassan09codechanges}. 
  \item Use  predictors to proactively fix defects~\cite{kamei16_lit, legoues12_aprlit, arcuri2011practical}
  \item Study defect prediction not only just release-level \cite{di18_fft, agrawal2018better} but also change-level or just-in-time \cite{yan18_tddetermination, kamei12_jit, nayrolles18_clever, commitguru} both for research and also industry. 
  \item Explore ``transfer learning'' where predictors from one project are applied to another~\cite{krishnaTSE18,nam18tse}.
  \item Explore the trade-offs between explanation and performance of defect prediction models~\cite{di18_fft}.
  \item Assess different learning methods for building models that predict software defects~\cite{ghotra15}. This has led to the development of hyperparameter optimization and better data harvesting tools \cite{agrawal2018wrong, agrawal2018better, Fu17easy, Fu16Grid, fu2016tuning,tantithamthavorn2016automated}. 
\ee
\noindent
The important thing to note about all these eight research areas is that all their conclusions are questionable if commit
messages are labelled incorrectly in the first place.
As \tbl{sample}
showed, ground truths of defect prediction can be inaccurate if the wrong labelling methods are applied. Hence, the concern here is that it could be misleading to train and draw conclusions from the wrongly labelled data.

\section{ Incremental Active Learning}
\subsection{Overview}

The case was made above that (a)~labelling commit messages is a vital task at the core of much current research, and (b)~manually labelling those commits is a very slow process. This
section describes the EMBLEM active learning method
that incrementally labelling a small subset of the commits. Using those labels, a machine learner can then find nearly all the remaining interesting/bug-fixing commits. Using these methods, humans have to read in detail only a small percentage of the commits (under 22\%, median value from the last column of \tbl{data}c). 

\begin{wrapfigure}{r}{1.4in} 
\includegraphics[width=1.4in]{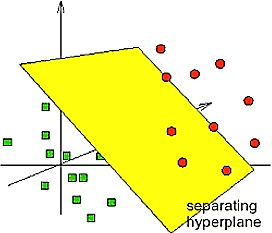}
\caption{ Separating bug-fixing (red) from non bug-fixing (green) files.}\label{fig:svm}
\end{wrapfigure}To understand active learning, consider the decision boundary between the bug-fixing commits and 
other commits shown in \fig{svm}. 
One tactic for quickly
finding those bug-fixing commits would be to ask humans to review and assess a few dozens of commits that fall into the red region of this figure, as far as possible from the green ones (i.e. {\em certainty sampling}). Another tactic would be to review items that are closest to the boundary (i.e. {\em uncertainty sampling}). Such {\em active learners} outperform supervised and semi-supervised learners and can significantly reduce the effort required to achieve high
recall in electronic discovery, evidence-based medicine, and reading research papers ~\cite{Cormack2017Navigating,Cormack2016Engineering,cormack2016scalability,cormack2015Autonomy,cormack2014evaluation,wallace2010semi,wallace2010active,wallace2011should,wallace2012class,wallace2013active,Yu2018,Yu2019}.

\subsection{Core Problem}

The core problem of labelling is how to find most of the relevant or bug-fixing commits with the least inspection of the development commits/logs. Formally, the general form (documented in \tbl{formal}) can be described
as following: start with the set of reviewed commits of $L = 0$ labels, prioritize which changes to be reviewed so as to maximize $|L_B|$ (bug-fixing commits) while minimizing $|L|$ (the total
number of assigned labels).

The central insight of this paper is that labelling textual data (e.g. development logs/commits) is analogous to reading research papers where there is a small fraction of target class (i.e. bug-fixing commits for labelling and relevant research papers for reading). Repeating successes of active learning in electronic discovery, evidence-based medicine, and reading research papers indicate its' effectiveness. The secret is that the active learning framework includes human assessors in the loop to investigate and find most of the bug-fixing commits while keeping the overall inspected commits minimal. Specifically, there are several differences: 
\bi
\item \textit{More informed learning method}: as \S2.1 indicated, labelling bug-fixing commits have been only utilizing keywords that may not be generalized to developers coming from different domains. EMBLEM does not assume the patterns when labelling the commits, but instead, it lets the human developers identify a few bug-fixing commits. Then it learns the discriminating patterns and efficiently found similar bug-fixing examples through text-mining pipeline (described in \S3.3). It is suggestive that our method can be generalized to various domains with no prior knowledge. 
\item \textit{Stopping mechanism}: Minimizing developers' effort is crucial, especially for the computational science domain where the efforts should be focused on improving state-of-the-art scientific theories. Therefore, our EMBLEM offers an incremental estimation of approximately how many more relevant or bug-fixing commits are left in the corpus (also described in \S3.3). This is useful since, without it, developers would not know if it is cost-effective to continue labelling.
\ei

\begin{table}[!t]
\footnotesize
\caption{Problem description for EMBLEM.}\label{tbl:formal} 
\label{tab: problem}
\begin{tabular}{ll}
\rowcolor{gray!10} 
$C$: & the set of all candidate commits (in the project).\\\rowcolor{gray!10} 
$B \subset C$: & the set of ground truth bug-fixing commits. \\\rowcolor{gray!10} 
$N = C \setminus B$: & the set of ground truth not bug-fixing commits.\\\rowcolor{gray!10} 
$L \subset C$: & the set of labelled/reviewed commits, \\\rowcolor{gray!10} & each review reveals whether a commit $x\in R$.\\\rowcolor{gray!10} 
$L_B = L\cap B$: & the identified bug-fixing (included) commits.\\\rowcolor{gray!10} 
$L_N = L\cap N$: & the identified none bug-fixing (excluded) commits.
\end{tabular} 
\end{table}

Hence, for our labelling problem,
we adopted an active learning framework developed for reading research papers in SE literatures~\cite{Yu2018,Yu2019,yu2019searching,yu2018total}, called EMBLEM. It has been proofed to be effective not only in reading research papers, but also in inspecting software security vulnerabilities~\cite{8883076}, finding self-admitted technical debt~\cite{fahid2019better}, and test case prioritization~\cite{yu2019terminator}. In this work, EMBLEM handles not only labelling commits to certify the previous approaches but also establishing the new ground-truth labels for future studies.

\subsection{Framework}

All those aforementioned tactics are built into EMBLEM~\cite{Yu2018,Yu2019},
the active learner used for this work.
When reading  commits, EMBLEM
initially uses uncertainty sampling to fast build a classification model (for bug-fixing
or non bug-fixing commit message), then switches to certainty sampling to greedily find bug-fixing commits. The machine learner (i.e. SVM) use this feedback from human to learn their models incrementally. These models are then used to sort the stream of commit messages such that humans read the most informative ones first
(and the commits are resorted each time a human offers a new label for a commit). From the nomenclature of \tbl{formal}, More specifically, EMBLEM executes as follows: \\
\begin{enumerate}[start=1,label={\bfseries Step \arabic*}]
\item \textbf{Feature Extraction:} Given a set of commits candidates $E$, EMBLEM extract features from each candidate as top $N_1=4000$ $L2$-normalized (the square root of the sum of the squared vector) terms with the highest 
TF-IDF\footnote{For token $t$, its tf-idf score: $\mathit{Tfidf}(t) = \sum_{d\in D} \mathit{Tfidf}(t,d)$, in which for token $t$ in commit or document $d$, $\mathit{Tfidf}(t, d)=w^t_d\times (\log \frac{|D|}{\sum_{d\in D} \mathit{sgn}(w^t_d)}+1)$ where $w^t_i$ is the number of times token $t$ appears in document $d$.} scores (after stop word removal). Initialize the set of labelled data points as $L \leftarrow \emptyset$ and the set of labelled positive data points as $L_B \leftarrow \emptyset$.
\item \textbf{Initial Sampling:}
EMBLEM starts by
randomly sampling unlabelled candidate studies until humans declare that they see $N_2=1$ relevant examples. In 
the context of this paper, ``relevant'' will
mean ``bug-fixing commit''.
\item \textbf{Uncertainty Sampling:}
Then, as human assessors offer labels, one example at a time, EMBLEM
start training and updating with weighting to control query with uncertainty sampling, until $N_3=30$ relevant examples are found.
Here, different weights are assigned to each class
($W_B= 1/|L_B|$, $W_N= 1/|L_N|$).
\item \textbf{Certainty Sampling:}
Next, EMBLEM trains further using
certainty sampling and 
Wallace's ``aggressive undersampling'' \cite{wallace2010semi}
that culls majority class examples closest to the decision boundary.
\item \textbf{Relevant Examples Estimation:}
EMBLEM
stops training when it is estimated that \mbox{$N_4=95\%$} of the relevant have
been found.
\end{enumerate} 
To generate the $N_4$ estimate,
whenever the SVM model is retrained, EMBLEM makes temporary ``guesses'' about 
the unlabelled examples (by running those examples through the classifier).
To turn these guesses into an estimate of the remaining bug-fixing commits, EMBLEM:\\
\begin{enumerate}
\item 
Builds a logistic regression model, using the guesses.
\item
Using that regression model, EMBLEM makes new guesses on the remaining unlabelled examples.
\item
Loops back to step1 until the new guesses are the same as the guesses in the previous loop.
\item
Uses this logistic regression model to estimate the remaining number of positive examples in the data. 
\end{enumerate}

The reader will note that there are many specific engineering decisions built into the above
design (e.g. the values  $\{N_1=4000, N_2=1, N_3=30, N_4=95\%\}$). Those decisions where made by 
Yu et al.~\cite{Yu2018} after exploring 32 different kinds of active learners.
They report that, using the above requirements,
EMBLEM
found more relevant items faster than the
previously reported state-of-the-art in incremental text mining
retrieval~\cite{wallace2010active, cormack2015Autonomy}. Further,
the $N_4$ estimator converged much faster and obtained better estimates
that other prominent estimators from the text mining literature~\cite{ros2017machine,Cormack2016Engineering,wallace2013active}.


Our pre-experimental belief was that EMBLEM
would require extensive tuning before it could be used for labelling Github commits. However, the effectiveness of EMBLEM was obtained
using Yu et al.'s original decisions \cite{Yu2018,Yu2019} without extensive tuning. Future improvements can be achieved by tuning different settings.

\input{data}

\section{Experimental Methods}

The rest of this paper performs the experiments
that explore the research questions shown in the introduction.
Note that, in all the following, when we say ``use EMBLEM'', than is shorthand for use EMBLEM until
95\% of the bug-fixing commit messages have been found (where that 95\% is estimated using the $N_4$ method discussed above).

\subsection{Data}

The case studies of this paper come from computational science software. It is important to study such software since that code    has a widespread social impact.
For example, weather forecasts generated from computational science
software can   predict the path of hurricanes. This, in turn,
allows (e.g.) effected home owners to better protect themselves from
damaging winds. 
For another example, computational science explores the properties
of new materials. Synthesizing new materials is very expensive so standard practice is to use software to determine
those properties (e.g. via a finite element analysis). This, in turn, enables (e.g.) the faster transition of new materials to industry. Moreover, more quality software would guarantee the computational science work more credible and more reproducible. Therefore, 
better software engineering improves computational
science software, which would lead to better (e.g.) weather predictions and the faster creation of new industries based on new materials.


Another important reason to study   computational science   is that CS can  be used to stress test
the generality of existing SE methods.
Consider the results of \tbl{sample} where a standard labelling method in SE (i.e. keywords to identify bug-fixing commits) failed very badly
when applied to CS software.
This result is suggestive (but not conclusive) evidence that (a)~prior work on analytics has {\em  over-fitted} methods (to systems like Apache);  and that (b)~there is {\em no better time} than now to develop new case studies (like CS).

Our  data was collected as follows.
Using our contacts in the computational science community
(from the Molecular Sciences Software Institute (MOLSSI), and the Science Gateways Community institute (SGCI)) we found
678 computational science  projects.
Researchers
warn against using all the Github data~\cite{bird09promise,agrawal2018we, eirini15promise, munaiah17curating} since
many of these projects are simple one-person prototypes.
Following  their advice, we applied the sanity checks of \tbl{data}b 
to select 59  projects with sufficient software development information. 
To create a manageable study, we selected ten of these at random (across a range of implementation
languages). One proved to have certain local corruptions so it was  dropped.
The remaining nine projects are listed in \tbl{data}a and \tbl{data}c. Specifically, the last column of \tbl{data}c indicating the percentage of the commits that are needed to be read by human through EMBLEM in order to identify the 95\% of the bug-fixing commits (i.e. 22\% as the median percentage).

The bug-fixing commits in these projects are then mapped to the previous code changes or commits that introduce the bugs.
For that purpose, we used    
Sliwerski, Zimmermann, and Zeller's SZZ's
algorithm~\cite{costa17szz, Kim08changes, Sliwerski05changes}
to work back in time to identify what code changes lead to the bug report
(and the SZZ implementation we used here came from Commit.Guru).
 Rodríguez-Perez et al.~\cite{RODRIGUEZPEREZ2018164}
report that at least 187 papers have also used SZZ
in the same way as these experiments of this paper.

For this analysis, we used Commit.Guru tool~\cite{commitguru}.
This tool was used since Commit.Guru supports web-scale data collection from on-line 
Github projects. Also, Commit.Guru labels commit using a set of keywords that are represented
standard practice in the commit-level defect prediction domain. \tbl{metrics}
shows the sensible information that we used from Github repositories through Commit.Guru \cite{mockus00changeskeys, kamei12_jit}. Note that these
features are language agnostic and divide almost equally into information about the code and information about how
humans are changing that code:
\bi
\item
One challenge data with multiple projects is
that they use different programming languages. Hence, it is best to use language-agnostic measures.
\item
Herslab~\cite{Tsay:2014} and Devanbu~\cite{Rahman:2013} argue convincingly
that it is important to collect
information about (a)~the humans changing
the code as (b)~the code itself.
\ei

\input{commitguru_features}


\subsection{Evaluation Criteria }

We choose not to evaluate defect predictors on any single criteria (e.g., not just  recall)
since succeeding in one criteria can damage another~\cite{fu2016tuning}. Also,
we deprecate the use of precision and accuracy since these can be misleading for data sets where the target class is somewhat
rare ~\cite{Menzies:2007prec}
(e.g. as shown in \tbl{data}c, four of our data sets have less than one-fifth buggy commits). 

Instead, we will evaluate our predictors on criteria
that aggregated multiple metrics, as follows:

{\small
\begin{equation}\label{eq:recall}
\mathit{Recall} = \frac{\mathit{TruePositives}}{\mathit{TruePositives} + \mathit{FalseNegatives}}
\end{equation}

\begin{equation}\label{eq:far}
\mathit{FalseAlarm Rate(FAR)} = \frac{\mathit{FalsePositive}}{\mathit{TruePositive} + \mathit{TrueNegative}}
\end{equation}

\begin{equation}\label{eq:g}
\mathit{G} = \frac{2 \cdot \mathit{Recall} \cdot \mathit{(1 - FAR)} }{\mathit{Recall} + (1 - \mathit{FAR})}
\end{equation}}

The G-score is the harmonic mean between recall and the compliment
of the false alarm rate. Hence, this value drops if
{\em either} the recall rate {\em or} the false alarm rate is high.
This scoring method is recommended for data sets like ours where some
of the test samples have imbalanced class distributions~\cite{shatnawi10g1,comments07}.

We also evaluate our results using the 20/80 rule from Ostrand et al.~\cite{ostrand05_predicting}. They say
a  ``good''
defect predictor selects the
 20\% of files  containing 80\% 
 of the defects 
 In the 
 literature,
 this 20/80 rule is often called $P_{opt}20$ (the percent of the bugs found after reading 20\%).
 $P_{opt}20$ is widely used in the literature and, for details on that measure, we refer the reader to those publications~\cite{menzies07dp, kamei12_jit, yang16effort, monden13cost, mende10effort, monden13cost, lo17_ifa, di18_fft}. For this paper,
 all we need to say about $P_{opt}20$ is the conclusions reached from this metric
are nearly the same as the conclusions reached via G-score. Note that for G-score and $P_{opt}20$, the {\em larger} values are {\em better}.

\subsection{Statistical Methods}\label{tion:stats}

This study ranks treatments using the Scott-Knott procedure recommended by Mittas \& Angelis in their 2013 IEEE TSE paper~\cite{mittas2013ranking}. This method
sorts results from different treatments, then splits them in order to maximize the expected value of differences in the observed performances
before and after divisions. For lists $l,m,n$ of size $\mathit{ls},\mathit{ms},\mathit{ns}$ where $l=m\cup n$, the ``best'' division maximizes $E(\Delta)$; i.e. the delta in the expected mean value before and after the spit: 

\[E(\Delta)=\frac{ms}{ls}abs(m.\mu - l.\mu)^2 + \frac{ns}{ls}abs(n.\mu - l.\mu)^2\]

Scott-Knott then checks if that ``best'' division is actually useful. To implement that check, Scott-Knott would apply some statistical hypothesis test $H$ to check if $m,n$ iare significantly different (and if so, Scott-Knott then recurses on each half of the ``best'' division). For this study, our hypothesis test $H$ was a conjunction of the A12 effect size test of and non-parametric bootstrap sampling; i.e. our Scott-Knott divided the data if {\em both} bootstrapping and an effect size test agreed that
the division was statistically significant (95\% confidence) and not a ``small'' effect ($A12 \ge 0.6$).

For a justification of the use of non-parametric bootstrapping, see Efron \& Tibshirani~\cite[p220-223]{efron94}. For a justification of the use of effect size tests see Kampenes~\cite{kampenes2007} who warn that even if a hypothesis test declares two populations to be ``significantly'' different, then that result is misleading if the ``effect size'' is very small. Hence, to assess the performance differences we first must rule out small effects. Vargha and Delaney's non-parametric $A12$ effect size test was endorsed by Arcuri and Briand~\cite{arcuri2011practical}.
This test 
explores two lists $M$ and $N$ of size $m$ and $n$ by computing the probability that numbers in one sample are bigger than in another as below:

\[
  A12 = \left(\sum_{x\in M, y \in N} \begin{array}{lr}
    1, & \mathit{if } x > y\\
    0.5, & \mathit{if } x == y
    \end{array}\right) / (mn)
\] \vspace{5pt}

\subsection{Learners Used in this Study}\label{tion:learners}
\vspace{5pt}
There are several approaches to build a defect predictor.
This paper
uses methods that are (a)~standard in the literature as well as some that
have (b)~recently shown much promise.
For a definition of ``standard in the literature'', we use the Ghotra et al.
ICSE paper that grouped 32 defect predictors into different ranks
(see Table 9 of~\cite{ghotra15}). For this study, we used Random Forests+J48 and Logistic Regression (which are two top-ranked 
learners, according to the Ghotra results).
Also, 
just for completeness, we use Support Vector Machines (which comes from their bottom rank).
For the other learners, we use one method reported very recently at FSE'19 
(FFTs, describe below) as well as a standard data imbalance correction algorithm called SMOTE.

\subsubsection{Logistic Regression (LR)}
Given a regression function $t$
   that combines many variables, Logistic Regresion maps $t$ into the range 0..1 using $u=1/(1+e^{-t})$~\cite{Witten:2011}. A binary classifier for class labels $x,y$ is then constructed using (e.g.)
$\mathit{if}\; u<0.5\; \mathit{then}\; x\; \mathit{else}\;y$.
    
\subsubsection{Tree Learners: J48 + Random Forests (RF)}
J48 recursively builds one decision tree by
finding the feature whose ranges most reduce {\em entropy} (which is a measure of the division of class labels that call into each range).
Using J48 as a sub-routine, our
Random Forests builds many trees,
each time using different subsets of the data rows $R$ and columns $C$\footnote{Specifically, using $\log_2{C}$ of the columns, selected at random.}. 
Test data is then passed across all $N$ trees and the conclusions are determined (say) a majority vote across all the trees~\cite{Breiman2001}. 

\subsubsection{Support Vector Machines (SVMs)}
SVMs created a hyperplane that maximizes the distance between the two classes to it to separate them (i.e., defective or not). 
    In this paper, following the results of Ghotra et al. \cite{ghotra15}, the Sequential Minimal Optimization
    (SMO) SVM technique is used. SMO analytically solves the large
    Quadratic Programming (QP) optimization problem which
    occurs in SVM training by dividing the problem into a series
    of possible QP problems~\cite{zeng2008fast}.

\begin{figure}[!b]
{\small
\begin{alltt}
  if  \hspace{16pt}   LA  \(<\)  10 \hspace{17pt}  then nonBuggy  
  else if   Entropy \(\le\) 0.65 then Buggy  
  else if   NS \(>\) 3 \hspace{22pt}    then Buggy  
  else if   FIX == 1    \hspace{18pt}then Buggy  
  else            \hspace{50pt}nonBuggy    
\end{alltt}}
\caption{An FFTs tree. Built using the features of \tbl{metrics}.
The first guard of that tree is {\em LA $<$ 10} where this predicts for a nonBuggy commit. If that guard is false, then the reasoning falls down to the rest of the tree.
}\label{fig:fft}
\end{figure}

\subsubsection{SMOTE}
    SMOTE is not a learner, but a data pre-processor.
    Given some data set where the number of positive and negative examples are not equal, SMOTE randomly discards members of the majority class while also creating synthetic examples of the minority class.
    For that creation, each row $x$ finds $y_1,..,y_5$ similar rows of the same class. It then picks of those rows 
    $y_i$ at random
    and creates a new example at a randomly selected distance between $x$ and $y_i$. Some recent results report that off-the-shelf SMOTE can be improved by some local tuning~\cite{agrawal2018better,bennin2018mahakil}.
We do not use such local tuning since recently is has been shown that such tunings
are out-performed by FFTs (see below).

\subsubsection{Fast and Frugal Trees (FFTs)}
All the  learners listed above execute in the same
    manner, no matter what evaluation criteria is used to assess their learned models.
    FFTs, on the other hand, change their reasoning based on the target evaluation criteria. 
    
    More specifically, in this study,
    FFTs change the way they rank numeric ranges according to the evaluation criteria. 
Specifically:
\bi
\item All numeric columns are divided by their medium values.
\item Each division is then sorted by the predicated goal, sorted best to worst (if the goal is changed, FFTs would change how it sorts the discretized ranges). For both goals, the evaluation criteria scores {\em higher} if {\em more} bugs are found.
\ei
This sort order is used as follows.
One FFT (in a FFTs model) is a decision tree made as a binary tree where all internal nodes
have one leaf node with a guard condition leading to a classification decision~\cite{martignon2008categorization}. Each leaf can have two guards; specifically either the range associated with least or most $E_i$.\pagebreak This means that,
for depth $d$, we build  $2^d$ trees.
For example,   for $d=4$, \fig{fft} shows one of the $2^d=16$ possible trees. FFTs build all 16 trees then sorts then using the G-score (by running the training data through each one). The best tree (as discovered on the training data) is then applied to the test data.

Despite the apparent simplicity of FFTs, Chen et al.~\cite{di18_fft}
reported that this method performs dramatically
better than many prior results
seen at recent ICSE conferences~\cite{ghotra15,agrawal2018better}.
(including those that used hyperparameter optimization
and data pre-processing with SMOTE). 
There are two possible
explanations for this superior performance. Firstly, due to their discretization policy, they can make better use of the evaluation criteria than any other learner (that builds
their models without reflecting over the evaluation criteria). Secondly, the FFTs training process (of selecting the best out of 16 possible models) is essentially an ensemble
learning algorithm (albeit a very simple one) and such algorithms have been known to perform better than solo learners~\cite{kocaguneli2012value}.

\subsection{Experimental Rig}\label{tion:eval}
For this study, it is important to test data not used in training (to avoids overfitting on the training and inappropriately inflating the test performance scores). To that end, we
exploited the software release structure of our projects. Specifically, if a project had $R$
releases, then our learners were trained on release $r$ and tested on release $r+1$.

\subsection{Ground Truth}

For this study to work, some ``ground truth'' must be accessed against which we can compare different methods for labelling commit messages and classification. That ground truth was generated as follows.

Using the power of pizza, we attracted half a dozen graduate students (computer science doctoral candidates) to spend a day labelling commit messages. Messages were labelled ``bug-fixing'' or ``none bug-fixing''
at a rate of 8 messages/minutes/person. If this seems fast, then
note that the median size of these commit messages is not large. All the messages of \tbl{sample}
fall within the 25th to 75th percentile of commit message size. As a sanity check, half of the labels were read by a second person. The observed disagreement rate was low, 14.6\%. Some of the examples of the conflict are listed below. The human labelers have disagreements where (1) attempts to fix might not guarantee an actual bug-fixing activity (e.g. RMG-PY, HOOMD, and PCMSOLVER examples) and (2) ``fixes'' that might not be quantified as true bug-fixing activities (e.g. AMBER and XENON). 
\bi
\item RMG-PY: ``Possible fix for the duplicate training reactions problem''. 
\item HOOMD: ``Attempt to fix external plugin builds''.
\item AMBER: ``Misc readline fix from S. Brozell''.
\item PCMSOLVER: ``Still debugging now i get the first compression, however, errors occur somewhere after that''.
\item XENON: ``PMD warning fixes in Torque integration test.''
\ei

At the end of the labelling session,
only 23,000 commits from four projects (out of a total of nine) had been labelled manually by human beings. The authors
of this paper considered reading on to manually label the remaining 22,000 commits from our other
five projects. But given the tedium of that process,
this was conjectured to introduce errors into our 
labels. Moreover, such manual methods are not appreciated in the industry. Hence, we tested if a faster semi-automatic method (i.e. EMBLEM) would suffice by investigating 100 of those labels (selected at random) from the rest five projects. These are manually cross-labelled 
to generate Table \ref{tbl:100}.
Note that our EMBLEM generated labels 
performed very well.





All this data was used as follows. In {\bf RQ1}, different labelling methods are compared to the ground truths (i.e. human labels) from the original four projects. During defect prediction of {\bf RQ2}, the human labels are utilized as the ground truths for
the original four projects. However, for the rest five projects, the defect predictor are trained on $p_{i}[keyword]$ and $p_{i}[EMBLEM]$ (where $p_i$ denotes
data from a project at version $i$) and in order to predict and test on $p_{i+1}[keyword]$ and $p_{i+1}[EMBLEM]$. In other words, the proceeding version's commits attributes (independent attributes, same for both $p_{i+1}[keyword]$ and $p_{i+1}[EMBLEM]$) and labels achieved from each labelling method (dependent metric) will be utilized to test the defect predictor trained on the current version's commits' attributes (independent metrics, same for both $p_{i}[keyword]$ and $p_{i}[EMBLEM]$) and respective label type (dependent metric). This method is applied and endorsed through previous studies that solely using automating keyword \cite{nayrolles18_clever, commitguru, kamei12_jit, catolino17_jitmobile}.

\begin{table}[!t]
\caption{ Labels generated by EMBLEM text classifier for 100 randomly selected commits. Scored via manual cross- inspection.
}\label{tbl:100} \vspace{-5pt}
\begin{center}
\begin{tabular}{ r|P{1.5cm}|P{1.8cm}}
 \multicolumn{1}{c|}{} & \multicolumn{1}{c|}{} & \multicolumn{1}{c}{False-Alarm}\\
 \multicolumn{1}{c|}{Dataset} & \multicolumn{1}{c|}{Recall} & \multicolumn{1}{c}{Rate}\\
\hline
PCMSOLVER & 72 & 21 \\
XENON & 96 & 21 \\
AMBER & 96 & 16 \\
HOOMD & 98 & 13 \\
RMG-PY & 89 &  3 \\
\end{tabular}
\end{center}
\end{table}


\section{Results}

{\bf RQ1: How close are EMBLEM and keyword labelling to human labels?}


\begin{table}[!b]
\caption{RQ1 results. Comparing EMBLEM vs Keywords (from Commit.Guru) for bug-fixing commits identification performance by comparing generated labels against ``ground truth''; i.e. those
labels assigned by human readers.}
\vspace{-5pt}
\label{tbl:rq1}
 
\begin{center}

\begin{tabular}{ r|P{1.1cm}|P{1.35cm}|P{1.1cm}|P{1.25cm}}
\multicolumn{1}{c|}{} & \multicolumn{2}{c|}{} & \multicolumn{2}{c}{False-Alarm}\\
 \multicolumn{1}{c|}{} & \multicolumn{2}{c|}{Recall} & \multicolumn{2}{c}{Rate}\\
\cline{2-5}
 Dataset & Keyword & EMBLEM & Keyword & EMBLEM \\
\hline
LIBMESH & 74\% & 97\% & 24\% & 17\% \\
ABINIT  &  81\% & 96\% & 29\% & 18\% \\
MDANALYSIS   & 84\% & 95\% & 42\% & 23\% \\
LAMMPS   & 81\% & 99\% & 73\% & 35\% \\
\end{tabular}
\end{center}
\end{table}

Software maintenance is a continuing process, and software developers have the domain expertise in understanding the commit logs. Human-labelled commit logs serve as the ground-truth labels. This question compares different labelling methods of standard SE keywords and our EMBLEM 
against the ground-truth labels. In Table \ref{tbl:rq1},
labels generated by humans are used to score labels proposed by EMBLEM or keywords. The results are apparent, EMBLEM can recall most of the labels that closest to the ground truth labels (up to 99\% with 96\% on median) while achieve lowest false-alarm rates (down to 17\% and 20\% on median). It is notable for the case of LAMMPS project when false-alarm rate from Keywords labelling went really wrong with 73\%. In conclusion, Keywords labelling result in lower recall and higher false alarms; i.e.:

\begin{RQ}{\normalsize{Compared to humans...}} 
EMBLEM was best at reproducing the ground truth (i.e. the human labels).
\end{RQ}

\begin{table}[!b]
\begin{center}
\caption{RQ2 results: data miners comparison. Win percentages of G-score (top) and $P_{opt}20$ (bottom). 
Gray cells highlight predicting methods that were top-ranked the most in that project
by the statistical tests of \tion{stats}. Each cell is in the format of $P (W/N)$ where W is the number of times one treatment won over the other, and N is the number of the releases per project, then
the percentage win P is calculated by $W/N$. (S= SMOTE, SVM= Support Vector Machine, RF= Random Forest, LR= Logistic Regression)}
\label{tbl:rq2_learners}
\footnotesize
\hspace{-8pt}\resizebox{1\linewidth}{!}{
\begin{tabular}{r@{~}|r@{~}|r@{~}|r@{~}|r@{~}}
\footnotesize
 & \multicolumn{4}{c}{\textbf{\% G-score Wins}} \\
\cline{2-5}
\begin{tabular}[c]{@{}c@{}} \textbf{Dataset} \end{tabular} & \begin{tabular}[c]{@{}c@{}} \textbf{SMOTE+RF}\end{tabular} & \textbf{SMOTE+SVM} & \begin{tabular}[c]{@{}c@{}} \textbf{SMOTE+LR}\end{tabular} & \textbf{FFTs} \\ \hline
AMBER & \cellcolor{gray!20} 67 (2/3) & 0 (0/3) & 0 (0/3) & 0 (0/3) \\ 
PCMSOLVER & 0 (0/1) & 0 (0/1) & \cellcolor{gray!20} 100 (1/1) & 0 (0/1) \\ 
RMG-PY & 0 (0/5) & 0 (0/5) & \cellcolor{gray!20} 40 (2/5) & 0 (0/5) \\ 
 HOOMD & 0 (0/5) & 0 (0/5) & 0 (0/5) & \cellcolor{gray!20} 40 (2/5)  \\ 
 LAMMPS & 0 (1/8) & 13 (1/8) & 0 (0/8) & \cellcolor{gray!20} 50 (4/8)  \\  
 ABINIT & 0 (0/8) & 0 (0/8) & 13 (1/8) & \cellcolor{gray!20} 50 (4/8) \\
 XENON & 17 (1/6) & 0 (0/6) & 17 (1/6) & \cellcolor{gray!20} 50 (3/6) \\ 
MDANALYSIS & 0 (0/7) & 0 (0/7) & 0 (0/7) & \cellcolor{gray!20} 72 (6/7) \\ 
LIBMESH & 0 (0/7) & 0 (0/7) & 0 (0/7) & \cellcolor{gray!20} 100 (7/7) \\
\end{tabular}}

\vspace{5mm}

\hspace{-8pt}\resizebox{\linewidth}{!}{ \begin{tabular}{r@{~}|r@{~}|r@{~}|r@{~}|r@{~}}
\multicolumn{1}{c|}{} & \multicolumn{4}{c}{\textbf{\% $P_{opt}20$ Wins}} \\
\cline{2-5}
\footnotesize
\begin{tabular}[c]{@{}c@{}} \textbf{Dataset} \end{tabular} & \begin{tabular}[c]{@{}c@{}}    \textbf{SMOTE+RF}\end{tabular} & \textbf{SMOTE+SVM} & \begin{tabular}[c]{@{}c@{}} \textbf{SMOTE+LR}\end{tabular} & \textbf{FFTs} \\ \hline
LAMMPS & \cellcolor{gray!20} 38 (3/8) & 13 (1/8) & 0 (0/8) & 0 (0/8)  \\  
XENON  & \cellcolor{gray!20} 33 (2/6) & 0 (0/6) & 17 (1/6) & 0 (0/6) \\ 
ABINIT & 0 (0/8) & \cellcolor{gray!20} 25 (2/8) & 0 (0/8) & 13 (1/8) \\ 
MDANALYSIS & \cellcolor{gray!20} 15 (1/7) & 0 (0/7) & 0 (0/7) & \cellcolor{gray!20} 15 (1/7) \\ 
AMBER  & 0 (0/3) & \cellcolor{gray!20} 33 (1/3) & 0 (0/3) & \cellcolor{gray!20} 33 (1/3) \\ 
HOOMD  & 20 (1/5) & 0 (0/5) & 0 (0/5) & \cellcolor{gray!20} 40 (2/5) \\ 
RMG-PY & 0 (0/5) & 0 (0/5) & 0 (0/5) & \cellcolor{gray!20} 40 (2/5) \\ 
LIBMESH & 15 (1/7) & 0 (0/7) & 0 (0/7) & \cellcolor{gray!20} 57 (4/7) \\ 
PCMSOLVER & 0 (0/1) & 0 (0/1) & 0 (0/1) & \cellcolor{gray!20} 100 (1/1) \\ 
\end{tabular}}
\end{center} 
\vspace{-5pt}
\end{table}
\begin{table}[!t]
\caption{RQ2 (top 2 tables). Win percentages of G-score (left) and $P_{opt}20$ (right). Gray cells highlight the labelling method that were top-ranked most in that project by the statistical tests of \tion{stats} (in $P (W/N)$ format). Treatments: Keyword+FFTs (K) and EMBLEM+FFTs (E)}
\label{tbl:rq2aaa}

\small 

\resizebox{\linewidth}{!}{
\hspace{-10pt}\begin{minipage}{0.54\linewidth}
\begin{tabular}{r@{~}|r@{~}|r@{~}}
\multicolumn{1}{c|}{} & \multicolumn{2}{c}{\textbf{\% G-score Wins}} \\
\cline{2-3}
\begin{tabular}[c]{@{}c@{}} \textbf{Dataset} \end{tabular} & 
\multicolumn{1}{c|}{\textbf{K}} & \multicolumn{1}{c}{\textbf{E}} \\ \hline
PCMSOLVER &  \cellcolor{gray!20} 100 (1/1) & 0 (0/1) \\ 
AMBER &  \cellcolor{gray!20} 67 (2/3) & 33 (1/3) \\ 
HOOMD & 40 (2/5) &  \cellcolor{gray!20} 60 (3/5)\\ 
RMG-PY  & 40 (2/5) &  \cellcolor{gray!20} 60 (3/5)\\ 
ABINIT & 25 (2/8) &  \cellcolor{gray!20} 63 (5/8) \\ 
LIBMESH & 28 (2/7) &  \cellcolor{gray!20} 72 (5/7)  \\  
MDANALYSIS & 28 (2/7) &  \cellcolor{gray!20} 72 (5/7) \\ 
LAMMPS & 25 (2/8) &  \cellcolor{gray!20} 75 (6/8)\\
XENON & 17 (1/6) &  \cellcolor{gray!20} 83 (5/6)  
\\  
\end{tabular}  %
\end{minipage} \
\hspace{25pt}\begin{minipage}{0.60\linewidth}

\begin{tabular}{r@{~}|r@{~}|r@{~}}
\multicolumn{1}{c|}{} & \multicolumn{2}{c}{\textbf{\% $P_{opt}20$ Wins}} \\
\cline{2-3}
\begin{tabular}[c]{@{}c@{}} 
\textbf{Dataset} \end{tabular} & \multicolumn{1}{c|}{\textbf{K}} & \multicolumn{1}{c}{\textbf{E}} \\ \hline
PCMSOLVER &  \cellcolor{gray!20} 100 (1/1) & 0 (0/1) \\ 
XENON &  \cellcolor{gray!20} 50 (3/6) &  \cellcolor{gray!20} 50 (3/6)\\
MDANALYSIS & 43 (3/7) &  \cellcolor{gray!20} 57 (4/7)\\ 
LIBMESH & 14 (1/7) &  \cellcolor{gray!20} 57 (4/7)\\  
HOOMD & 40 (2/5) &  \cellcolor{gray!20} 60 (3/5)\\ 
LAMMPS & 25 (2/8) &  \cellcolor{gray!20} 63 (5/8)\\
ABINIT & 25 (2/8) &   \cellcolor{gray!20} 63 (5/8)\\
AMBER & 33 (1/3) &  \cellcolor{gray!20}  67 (2/3)\\
RMG-PY & 0 (0/5) &  \cellcolor{gray!20}  80 (4/5)
\\
\end{tabular}
\end{minipage}}
\vspace{10pt}
\end{table}

\textbf{RQ2: Does keyword labelling lead to better predictors for buggy commits?}

{\bf RQ2} 
checks how well those generated labels (from {\bf RQ2}) predict for defects. Again, from labels for bug-fixing and not bug-fixing commits, we applied
the SZZ algorithm from Commit.Guru to find which code commits that lead to bugs.
All such commits are labelled ``buggy=yes'' and others are labelled ``buggy=no'' where buggy means bug-inducing.

For this research question, we first experiment with different learning approaches that are standard in defect prediction literature including FFTs, Logistic Regression, Random Forests, Support Vector Machines, and the SMOTE preprocessor. The data here is labelled by EMBLEM. \tbl{rq2_learners} shows those results, where gray cells denote treatments with superior performance. FFTs performs as well or better in the majority cases (6/9 projects) for both G-score and $P_{opt}20$. FFTs built the best classifiers for buggy commits. \pagebreak

As endorsed by the previous experiment, FFTs is picked to test the effectiveness of the labelling method (keywords or EMBLEM). \tbl{rq2aaa} compares predictive performance using FFTs+EMBLEM and
FFTs+standard keyword labelling (using Commit.Guru). As before, the treatments with more superior performance are denoted by gray color.
Note that, in the majority case for the top two tables for Computational Science projects (7 out of 9 projects for both G-score and $P_{opt}20$), FFTs+EMBLEM performs best. That is:

\begin{RQ}{\normalsize{Compared to keyword labelling...}} 
EMBLEM generated better predictors for buggy commits.
\end{RQ}

{\bf RQ3: How much effort is saved by EMBLEM? }

Through our experiences in manual labelling the above studies, we offer the observations
of Table \ref{tbl:time1} and \ref{tbl:time2}.
These tables detail the time and cost required
to complete our work. Using this information, we can now
justify the calculations from \tion{iia}.
Recall that those calculations based on the aforementioned assumptions from \tion{iia}. These showed that for large
labelling tasks, the methods of this paper can reduce the
resources required for labelling by over an order of magnitude ($\approx$ eight times). As seen in Table~\ref{tbl:time2}, EMBLEM saved from manual labelling for both money and time from \$320K to just under \$40K and from 39,000 hours to 4,940 hours. This result is achieved from committed to finding 95\% of bug-fixing commits from only reading 22\% of the commits corpus. 

\begin{RQ}{Compared to manual labelling...} 
EMBLEM-based labelling can be an order of magnitude cheaper.
\end{RQ}


\section{Why Does It Work?}

Stepping back from the specifics of the above quantitative results, this section offers a more qualitative reflection on these results.
Here we ask: what is the core advantage of EMBLEM? What is the explanation for its superior performance?

\begin{table}[!t]
\caption{RQ3 results: time cost model for manual versus EMBLEM labelling}
\vspace{-10pt}
\label{tbl:time1}
 
\begin{center}
\begin{tabular}{ l|c|c }
 \multicolumn{1}{c|}{} & \multicolumn{1}{c|}{Manual} & \multicolumn{1}{c}{EMBLEM}\\
\hline
time per commit & 7 secs & 4 secs \\
commits need to read & 5000 & 1100 \\
time per project & 19.44 hrs & 2.47 hrs \\
time per project (with \S2.2 assumptions) & 78 hrs & 9.88 hrs \\
time for 500 projects & 39,000 hrs & 4,940 hrs \\ 
\end{tabular}
\end{center}
\end{table}

\begin{table}[!t]
\caption{RQ3 results: money cost model for manual versus EMBLEM labelling}
\label{tbl:time2}
\vspace{-5pt}
\begin{center}

\begin{tabular}{ l|c|c }
 \multicolumn{1}{c|}{} & \multicolumn{1}{c|}{Manual} & \multicolumn{1}{c}{EMBLEM}\\
\hline
commits per hour & 514 & 900 \\
money per project & \$160 & \$20 \\
money per project (with \S2.2 assumptions) & \$640 & \$80 \\
money for 500 projects & \$320,000 & \$40,000 \\ 
\end{tabular}
\end{center}
\vspace{2mm}
\end{table}

In reply, we conjecture that much of the current work on empirical software might have explored a very narrow range of tools.
We say this since our reading of the literature is that most of that work has focused on artifacts taken from the Microsoft or Google universe (or by hobbyist developers working on open-source tools that handle
general tasks like databases or GUI design). Clearly, there is much benefit in exploring those tools since, after all, they are widely used with high impacts (e.g. explore large scale phenomena of hurricanes).
On the other hand,
one of the powerful benefits of software engineering is that it can be customized to any number of application niches.
For example, much of the data explored here comes from the computational physics community. We speculate that:

\bi
\item
In that application niche of computational physics,
the language used by developers (who are physicists) is different from the language used by Google/Microsoft/open source developers.
\item
Hence, EMBLEM does better than standard keywords for this application niche since, using EMBLEM, it is possible to learn the nuances of
that language. 
\ei

For specific examples 
of this specialized kind of language, please consider the words found most interesting (by TF-IDF) in the  RMG-PY project:
\bi
\item The five most important words were ``molecule'', ``reaction'', ``kinetic'', ``thermo'', and ``mechanism''.
\item
Note that these terms are more specialized to the 
physics domain than to general software engineering.
\ei

Other researchers also note that the language used in niche application
areas (such as computational physics) can be different from general SE. Carver, Bieman, and Tu et al. \cite{bieman18_slr, carver07_environment, tu2020changing}
indicate that scientific projects' development and verification/bug-fixing activities are mostly based on exploring
scientific issues which often are ``too complex, too large, too small, too dangerous, or too expensive to explore in the real
world'' \cite{segal08_ss}. \revised{Specifically, a Github issue from the deal.II finite element library\footnote{deal.II: computational solution of partial differential equations using adaptive finite elements \cite{BangerthHartmannKanschat2007} (more than 1250+ citations).} (\#8937) described their issue of having the system froze with 4+ billion degrees of freedom and no convergence of the linear solver. Therefore, it can be very different than performing testing at Microsoft and Google. It requires domain experts to look extensively into the simulation, the output, the graph, etc. Hence, EMBLEM is the first step toward a specialized approach to incorporate that human expertise to better SE for domain CS.}

The main message here is that:
\begin{quote}
{\em Different tools have to be tuned in different ways for different domains.}
\end{quote}

Of course, that conclusion could have been reached prior to all the work of this paper. This study serves as a proof of concept of the message above for tailored SE tools for CS domain. It
demonstrates how that tuning process (which requires the manual inspection of many arguments) can be optimized by active learning.
In the particular case of defect prediction on commit level, we showcased that optimization can be as much as an order of magnitude faster (approximately eight times, see our {\bf RQ3} results).

\section{Threats to Validity }

\subsection{Evaluation Bias}

This paper employed the G-score as defined in Equation 3.
This value is the harmonic mean between recall and false-alarm of risky software commit prediction power. There are other evaluation scores that could be applied to this kind of analysis~\cite{lo17_ifa}
and, in the future, it would be useful to test in the central claim of this paper holds for more than just G-scores and $P_{opt}20$.

\subsection{Learner Bias}

This study utilized FFTs and compared it with 
Logistic Regression, Random Forest, and Support Vector Machine (in the combination with SMOTE). 
The case was made in \S4.4 that this represents an interesting range of current practice.
Nevertheless, it might be useful in future work to test if the central claim of this paper (that a combination of human+artificial intelligence, called EMBLEM, is a good way to label commit messages)
hold across multiple classifiers.

\subsection{Sampling Bias}

Like any data mining paper,
our work is threatened by sampling bias; i.e. what holds for the data we studied here may
not hold for other kinds of data. 
Within the space of one paper, it is hard to avoid sampling bias.
However, what researchers can do is make all their scripts and data available
such that other researchers can test their conclusions whenever new data becomes available. To that end, we have made all our scripts and data available at github.com/sillywalk/defect-prediction/.

That said, and to repeat the message of the last section, when different methods work for different data, researchers must take the time to carefully check the ground truth in the new data. Let that check be overwhelmingly slow and expensive, we recommend the use of active learning with tools like EMBLEM.

\subsection{Construct Validity}

We acknowledge that our work still requires a portion of the data to be labelled which is different from existing techniques that require full access to the data. The bigger the bug-fixing commits portion within software development, the more the expert have to label
(as demonstrated in Table 5c and Table \ref{tbl:diff}). This
means
that our work would be most beneficial in the space where the target label only makes up of a minority portion of the whole data. Such prior work mentioned in \S2.2 would be benefit via the application
of   EMBLEM.
\revised{Having an expert to read 23\% of the commits does save up to \textbf{eight} times on average in comparison to manual labelling. Moreover, our framework allows generalization for other domains outside of computational science and can be utilized to study tasks that only make up a minority portion of the space. }

\revised{Beyond the results
of this paper, it is of course
desirable to further reduce 
the amount of  
commits that need to be read.
That said, if the amount read
by humans starts approaching
0\%, then  we    risk losing the
value-added nature of human expertise.
We recommend that further work in this 
area looks to decrease, but not remove,
human involvement in the labelling process.
For example,
future work could   clustering together
similar projects and only label a smaller portion of each clusters.
Such a {\em semi-supervised}
approach is similar to Lo et al.'s work that has explored only Linux software \cite{Tian12_linux}.}

\begin{table}[!t]
\vspace{5pt}
\caption{Checking external validity. Results when replicating the experiment from RQ2 for 2 standard SE projects. Win percentages of G-score (left) and $P_{opt}20$ (right). Gray cells highlight the labelling method that were top-ranked most in that project by the statistical tests of \tion{stats} (in $P (W/N)$ format). Treatments: Keyword+FFTs (K) and EMBLEM+FFTs (E)}
\label{tbl:rq3}
\vspace{10pt}
\footnotesize 
\resizebox{\linewidth}{!}{
\begin{minipage}{0.55\linewidth}
\begin{tabular}{r@{~}|r@{~}|r@{~}}
\multicolumn{1}{c|}{} & \multicolumn{2}{c}{\textbf{\% G-score Wins}} \\ [0.05cm]
\cline{2-3}
 \begin{tabular}[c]{@{}c@{}} \textbf{Dataset} \end{tabular} & 
\multicolumn{1}{c|}{\textbf{K}} & \multicolumn{1}{c}{\textbf{E}} \\ [0.05cm]
\hline
E.PLATFORM & 50 (4/8) & 50 (4/8) \\ [0.1cm]
E.JDT &  14 (1/7) & \cellcolor{gray!20} 71 (5/7) \\ 
\end{tabular}
\end{minipage}
\begin{minipage}{0.55\linewidth}
\begin{tabular}{r@{~}|r@{~}|r@{~}}
\multicolumn{1}{c|}{} & \multicolumn{2}{c}{\textbf{\% $P_{opt}20$ Wins}} \\
\cline{2-3}
\begin{tabular}[c]{@{}c@{}} \textbf{Dataset} \end{tabular} & 
\multicolumn{1}{c|}{\textbf{K}} & \multicolumn{1}{c}{\textbf{E}} \\ 
\hline
E.JDT &  \cellcolor{gray!20} 71 (5/7) & 29 (2/7) \\  [0.1cm]
E.PLATFORM &  \cellcolor{gray!20} 62 (5/8) & 38 (3/8) \\ 
\end{tabular}
\end{minipage}}
\vspace{5pt}
\end{table}

\subsection{External Validity}


{\bf RQ1, RQ2} demonstrated EMBLEM's effectiveness over keyword for computational science projects while {\bf RQ3} showcased how efficient EMBLEM is over manual labelling.
But is EMBLEM's competency a factor of our test cases? Or does it work for more than just the
computational science projects studied here?
To test that, we applied EMBLEM to a sample of standard large-scale SE projects, Eclipse JDT (E.JDT) and Eclipse Platform (E.PLATFORM). We used this data since it has previously been used in
several prior papers \cite{kamei12_jit, yang2015deep}.

\begin{table}[!t]
\vspace{7pt}
\caption{Distribution of commit message size through a number of words in Median and Standard Deviation of computational science projects (top chart) and from standard SE projects (bottom chart). Gray cells denote median values for each statistical measure. }\label{tbl:text_dist}
\begin{center}
\begin{tabular}{ r|P{1.5cm}|P{1.8cm}}
 \multicolumn{1}{c|}{} & \multicolumn{1}{c|}{} & \multicolumn{1}{c}{Standard} \\
 \multicolumn{1}{c|}{Dataset} &
 \multicolumn{1}{c|}{Median} & \multicolumn{1}{c}{Deviation}\\
\hline
ABINIT & 29 & 117 \\
XENON & 	37 & 	31\\
OPENMM & 	38 & 	46\\
HOOMD & 	47 & \cellcolor{gray!20}	74\\
AMBER & 	\cellcolor{gray!20} 50 & 	143\\
MDANALYSIS & 	51 & 	119  \\
LAMMPS & 	53 & 8 \\
LIBMESH & 	74 &  70 \\
RMG-PY & 	83 & 108 \\
\end{tabular}

\vspace{0.75cm}

\begin{tabular}{r|P{1.5cm}|P{1.8cm}}
 \multicolumn{1}{c|}{} & \multicolumn{1}{c|}{} & \multicolumn{1}{c}{Standard} \\
 \multicolumn{1}{c|}{Dataset} &
 \multicolumn{1}{c|}{Median} & \multicolumn{1}{c}{Deviation}\\
\hline
E.JDT  &  17  &  69  \\
E.PLATFORM  &  19  &  1526 \\
TRAFFICCONTROL & \cellcolor{gray!20} 34 & 86 \\
TENSORFLOW & 64 & 360\\
LUCENE & 92 & \cellcolor{gray!20} 264\\

\end{tabular}
\end{center}
\end{table}

As shown in Table \ref{tbl:rq3}, assessed in terms of $P_{opt}20$, we would prefer EMBLEM's labelling methods. In that table, we see that EMBLEM performs the same or much better than using standard keywords. This is an important result since $P_{opt}20$ addresses the core business case for defect prediction. As argued by Arisholm, and Briand~\cite{Arisholm:2006}
and Ostrand et al.~\cite{ostrand05fault}, predictors with high  $P_{opt}20$ are
better since they mean developers waste less time looking for
bugs in the code predicted to be faulty.

On the other hand, as shown in Table \ref{tbl:rq3}, assessed in terms of G-Score, EMBLEM performs worse
than using keywords for these two data sets.
On investigation, we found that this is due to two rather extreme quirks
of the data used in that study that is described below. We took extra steps to investigate the existence of the quirks in a few 
standard SE projects on Github including TRAFFICCONTROL, LUCENE, and TENSORFLOW. 
It is notable that those quirks are highly dependent on the nature
of those datasets-- so much that it would be useful to have ready access to alternate labelling methods such as EMBLEM. 

The first data quirk we found was that the format of the commits from SE community is very standardized. In the data of Table \ref{tbl:rq3}, over half of the commit comments were imitated some variation of the form
``\textit{Bug X: [type\_of\_bug] description\_of\_the\_bug-fixing\_activity}''. This format is perhaps not surprising for well-known projects in the SE community. For example, TRAFFICCONTROL's commits are formatted as ``\textit{[activity\_type]: description (\#\_reference\_to\_Github\_issue)}''. Moreover, commits from most projects within the Apache system on Github (i.e. LUCENE) are formatted with: ``\textit{[name\_of\_the\_project]: description\_of\_the\_commit}''.

This suggests that these comments were actually written by another
system, which then used Github as a storage facility. Hence we say, that if a project
has a highly stylized and rigid commit comment structure, then keywords
are a natural choice for labelling. However, for other kinds of ``wilder'' projects, where developers
have different backgrounds outside of standard SE (e.g. certain designs and specifications of a processor) with more freedom in their documentation, a different method would be more appropriate. Specifically, Table \ref{tbl:text_dist}, the median length of commit messages from computational projects are frequently 1.5 (50/34) times more than standard SE projects. However, the standard deviation attribute indicates a different story. Even with higher commit message size, the commit message size 
values from computational science projects are more tightly
compact around the median values in comparison with the commit message size values of 74 while for standard SE projects it is around 264. Therefore, for these ``wilder'' projects, then EMBLEM is recommended.

\begin{table}[!t]
\vspace{7pt}
\caption{Bug-fixing commits rate from computational science projects (top chart) and from standard SE projects (bottom chart). Gray cells denote median values.}\label{tbl:diff}
\begin{center}
\begin{tabular}{ r|P{1.8cm}}
 \multicolumn{1}{c|}{} & \multicolumn{1}{c}{Bug-fixing} \\
 \multicolumn{1}{c|}{Dataset} & \multicolumn{1}{c}{Commit Rate} \\
\hline
LAMMPS & 11\% \\
PCMSOLVER & 13\% \\
RMG-PY & 16\% \\
ABINIT & 18\% \\
LIBMESH & \cellcolor{gray!20} 23\% \\
HOOMD & 25\% \\
AMBER & 27\% \\
MDANALYSIS & 29\% \\
XENON & 41\% \\

\end{tabular}

\vspace{0.75cm}

\begin{tabular}{r|P{1.5cm}}
 \multicolumn{1}{c|}{} & \multicolumn{1}{c}{Bug-fixing}\\
 \multicolumn{1}{c|}{Dataset} & \multicolumn{1}{c}{Commit Rate} \\
\hline
TRAFFICCONTROL  & 54\% \\
LUCENE & 59\% \\ 
TENSORFLOW &  \cellcolor{gray!20} 61\% \\
E.PLATFORM & 73\% \\
E.JDT & 85\% \\
\end{tabular}
\end{center}
\vspace{-20pt}
\end{table}

The second data quirk is shown in \tbl{diff}. This table contrasts the types of commits
seen in the data sets of this paper. Note that the amount of bug-fixing commits seen in SE projects (61\%)
are almost three times more likely than the median values seen in the computational science software development (23\%).
This suggests that these SE projects are mostly in maintenance mode (where most of the development are concerned with preserving the software) while the scientific software is more in development mode (where new features are added frequently). That is,  
for highly mature projects (such as these SE ones)
where:
\bi
\item
Little new functionality is being added;
\item
Much activities are already defined and standardized; 
\ei

Therefore, simple
keywords are a natural method for labelling bug-fixing commits.
In such highly standardized projects, it is convenient and possible to catch bug-fixing activities through a few keywords (e.g. bug, fix, error, etc). 
However, for more dynamic projects,
where much functionality is being changed continuously, such active-learning frameworks like EMBLEM that can better adapt to new situations
to develop
local language models for the current context of the project.

In summary, and in support of the general theme of this paper, this external validity experiment demonstrates the danger of treating
all data with the same method (e.g. keywords list of Table~\ref{tab:words}). When new data arrives, it is essential to incrementally review it and verify if old conclusions hold for the new data. That verification can be time-consuming, labor-intensive, and costly unless some kind of active learning assistant is applied to suggest what examples are best to examine next. Hence, EMBLEM is recommended for checking and labelling the data.

\pagebreak



\section{Future Work}
As for future work, there are many options.
For example, 
we could repeat this study on more data or a different domain.

Also,
we could explore other control parameters for EMBLEM.
All the above results were obtained using Yu et al.'s \cite{Yu2018,Yu2019} original requirements (e.g. the values of $\{N_1=4000, N_2=1, N_3=30, N_4=95\%\}$) within the EMBLEM method. It is possible that other settings for these parameters could lead to better results.

Further,
all the work studied here relied on an off the-shelf Sliwerski, Zimmermann, and Zeller's SZZ's
algorithm~\cite{costa17szz, Kim08changes, Sliwerski05changes} (implemented through Commit.Guru \cite{commitguru})
that traced back to locate the bug inducing commits.
SZZ itself is an algorithm under active research
and there are numerous proposed improvements \cite{costa17szz}
that could be useful for our work.

Another issue is that all the data-mining process in this paper focused on learning the change level or commit-level as a whole. 
In this approach, changes attributes on multiple files (from Table \ref{tbl:metrics}) are averaged out within a commit \cite{kamei12_jit, commitguru, nayrolles18_clever, yan18_tddetermination}. This tend to be language agnostic and the results can be generalized. 
However, that approach might be enhanced by including features extracted within the commit messages
(as proposed by Yan et al. \cite{yan18_tddetermination})
or by including file or function attributes (i.e. learning on static code attributes such as C.K. and McGabe metrics) \cite{dpmetrics94, fu2016tuning, danijel13lit, Jureczko10oodp, ghotra15, nagappan05, menzies07dp, menzies10dp, krishna16bellwether, nam18tse, agrawal2019dodge, di18_fft, Tu19_hyper, agrawal2018better} that are more granulated and high-dimensional.

Finally, data mining technology keeps evolving.
Agrawal et al. \cite{agrawal2019dodge} recently argued
that for any dataset where FFTs are effective, that there is a better
algorithm (that they call 
$DODGE(\epsilon)$). Moreover, Yang et al. \cite{yang2015deep} designed the deep belief network to generate more quality metrics from the given metrics by Kamei et al. \cite{kamei12_jit}. Both are promising avenues for future work.

\section{Conclusion}

From bug-fixing or not bug-fixing labels, the ground-truth of software bugs (serve as the core for many defect-related analytics work) is obtained. 
However, different kinds of software used 
different language to describe their bugs. Hence, as shown in \tbl{rq2aaa}, standard labelling methods can perform badly when applied to new kinds of software (e.g. the computational science projects explored here).

Intuitively, one way to find the labels is to create teams of humans to manually read all the commits. As details in \tion{iia}, that process can get very expensive while receiving very little attention. Note that much research in labelling assumes the existence of pre-labelled datasets that are still costly to curate and certify. 

The standard approach is to use a set of defined keywords to capture the characteristics of the commit logs which might not be able to generalize to other diverse domains outside of traditional software engineering because (1) not every commit logs will include words such as ``bug'' and ``fix'' and (2) the semantics of these commit logs might focus more on the scientific theories instead of the implementation details (as discussed in \S6). 

Therefore, a more efficient way to find and certify the labels is to use incremental AI tools that learn an
appropriate local model. Such AI tools can present
examples to a human, one at a time. Whenever a human
offers a label to an example, the AI can update its internal model.
This internal model can be used to look ahead to find the next most-likely-to-be-bug-fixing example.
After a few loops of this process, the AI tool might be able to learn a model
that can find nearly all the remaining bug-fixing commits (95\% of them).

This paper has applied and evaluated one such AI+human partnership method, an incremental SVM method called
EMBLEM. On experimentation, we found that:
\bi
  \item At the labelling level, with human-labelled as ground-truths, EMBLEM maximizes recall and minimizes false-alarm for buggy commits identification more than automatic Keyword tagging method (see {\bf RQ1}). 
  \item Moreover, EMBLEM provides higher quality data for better performance of buggy commits prediction model (i.e. FFTs) than Keyword 78\% of the cases study here when evaluating with G-score and $P_{opt}20$ (see {\bf RQ2}).
  \item Even if EMBLEM is not more effective to label standard SE projects development than the keyword system, we can still recommend this system. EMBLEM can reduce the time required to import and label commit messages by more than an order of magnitude ($\approx$ eight times cheaper, see {\bf RQ3}). 
\ei
We note that quality and efficient labelling is not only needed in defect prediction but also a wide range of research tasks (e.g see the list \S 2.3). While this study has only examined defect prediction,
we see no reason why its message and methods should not hold for other kinds of analytics. In future work, we plan to apply the methods of this paper to other domains.

Finally, this paper also demonstrated and attempted to address the general area of data sharing and research reproducibility in software engineering (see \S 2.1). It is our hope that this work will lay down a foundation and motivate further in-depth investigation on labelling specifically and better research practices in software engineering generally.


\section*{Acknowledgements}
This work was partially funded by an NSF CISE Grant \#1826574 and \#1931425.

\bibliographystyle{plain}
 \bibliography{main.bib}
\newpage
\begin{minipage}{.45\textwidth}

\begin{IEEEbiography}[{\includegraphics[width=1.05in, height=1.4in,clip,keepaspectratio]{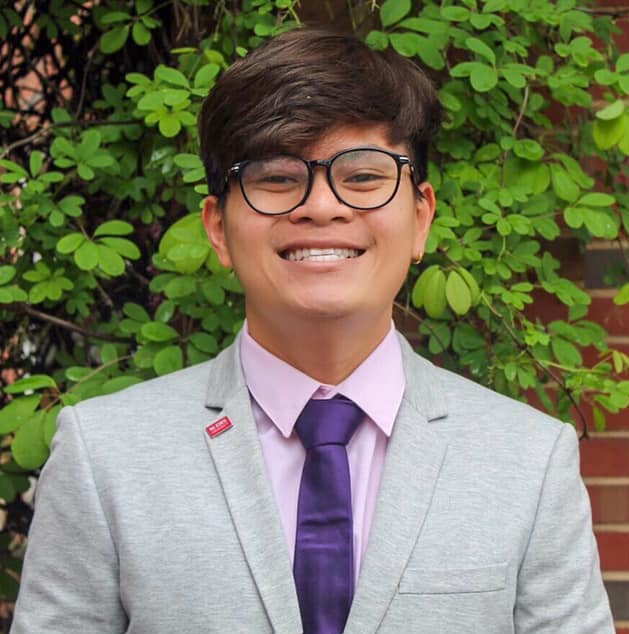}}]{Huy Tu} is a third year Ph.D. student in the department of Computer Science at North Carolina State University. He explores machine learning models that support and leverage from the human experience to solve real-world problems in software engineering. For more information, please visit \url{http://kentu.us}.
\end{IEEEbiography}
\begin{IEEEbiography}[{\includegraphics[width=1in,clip,keepaspectratio]{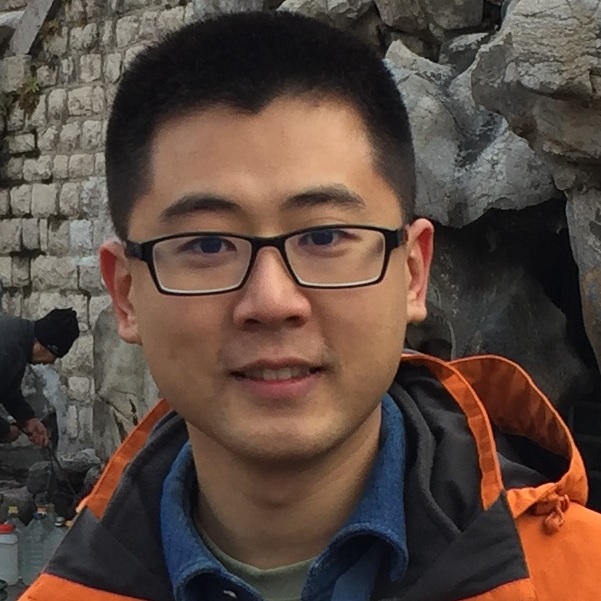}}]{Zhe Yu}
 (Ph.D. NC State, 2020)  is an assistant professor in software engineering at the  Rochester Institute of Technology, USA,  where he teaches data mining and software engineering. His     research explores  collaborations of human and machine learning algorithms that leads to better performance and higher efficiency.  For more information,  please visit \url{http://azhe825.github.io/}
\end{IEEEbiography}

\begin{IEEEbiography}[{\includegraphics[width=1.05in,clip,keepaspectratio]{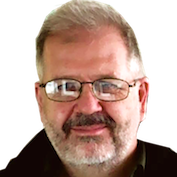}}]{Tim Menzies} (IEEE Fellow, Ph.D. UNSW, 1995)
is a Professor in comptuer science  at NC State University, USA,  
where he teaches software engineering,
automated software engineering,
and programming languages.
His research interests include software engineering (SE), data mining, artificial intelligence, and search-based SE, open access science. 
For more information,  please visit \url{http://menzies.us}.
\end{IEEEbiography}
\end{minipage}
\end{document}

%% file: sample_commits.tex
\begin{table*}[!t]
\caption{Commit messages from computational
software systems (see~\tbl{data}). Each commit is labeled ``bug-fixing'' by either
a keyword method (from Commit.Guru\cite{commitguru})) or Emblem. 
Right-hand side comments come from a manual inspection of each commit.} 
\scriptsize
\begin{center}
\vspace{-5pt}
\hspace{-5pt}\resizebox{1\linewidth}{!}{
\begin{tabular}{l|cc|c}
 &  \multicolumn{2}{|c|}{\small{Label=bug-fixing?}} & \small{Comment on the}\\

\small{Commit message} & \small{Keyword} & \small{EMBLEM} & \small{Keyword labels}\\ 
\hline
\texttt{fixes \#143: alignto() now checks the 2 selections describe the same atom} & Y &	 Y &   \\ 
\texttt{Correct bugs due to merge (rhotoxc)}	& Y &	 Y & \small{Correct}\\ 
\texttt{Convert tsmear to tphysel in vtorhotf.F90} &	 N &	 N &  \\ 
\texttt{Universe can load multiple trajectories from positional args} &	 N &	 N &   \\ \hline
\texttt{NetCDFWriter working (closes Issue 109)}	& N&	 Y &  \\ 
\texttt{Correction in magnetization rotation (DFPT+PAW)} &	 N &	 Y &  \\
\texttt{Add missing module dependency} &	 N &	 Y & \small{False-Negative}\\ 
\texttt{findSubgraphIsomorphisms works if you pass it a complete mapping} &	 N &	 Y &  \\ 
\hline
\texttt{documentation updates and fixes} &	 Y &	 N &   \\ 
\texttt{Removed unused expected error from Selections} &	 Y &	 N &   \\ 
\texttt{Test for HOLE changed form error to warning when HOLE binary is not there} &	 Y &	 N & \small{False-Positive}  \\ 
\texttt{Added CHANGELOG entry for fix of Issue \#550.} &	 Y &	 N &  \\ 
\hline
\end{tabular}}
\label{tbl:sample}
\end{center}

\end{table*}

%% file: data.tex
\begin{table*}
\caption{Data used in this study.}\label{tbl:data}
\vspace{-5pt}
~\hrule~
\vspace{1.5mm}

\begin{minipage}{.45\linewidth}
{\bf     \tbl{data}a: Data selection and pruning.}


678 computational science projects were identified.
Many of these are purely personnel projects (or just used for web storage)
so following the advice of Kalliamvakou et al.~\cite{Kalliamvakou:2014}, we used the sanity
checks of     \tbl{data}b to prune these to 59 projects.
On these, we use nine (selected to cover a range of languages):
\bi
\item PCMSOLVER:  API to the Polarizable Continuum Model \cite{pcmsolver}.
\item XENON: middleware interface to compute \& storage resources \cite{xenon}.

\item MDANALYSIS: Python code to analyze molecular dynamics trajectories generated from other simulation packages \cite{mdanalysis}.
\item HOOMD: particle simulation for hard particle Monte Carlo simulations of a many shape classes \cite{hoomd}. 
\item ABINIT: an atomic-scale simulation software suite \cite{Abinit}.
\item AMBER: Fast, parallelized molecular dynamics analysis \cite{Amber-MD}.

\item RMG-PY: the Python Reaction Mechanism Generator. Generates chemical reaction mechanisms for modeling reaction systems \cite{ReactionMechanismGenerator}.
\item LAMMPS: Large-scale Atomic/Molecular Massively Parallel Simulator, a classical molecular dynamics simulation code \cite{lammps-sandia}.
\item LIBMESH:  numerical simulation of partial differential equations on serial and parallel platforms. \cite{libMesh}.
\ei
For statistics on these systems, see     \tbl{data}c.

\end{minipage}~~~~\begin{minipage}{.55\linewidth}
\begin{center}
{\bf  \tbl{data}b: Sanity checks (designed using~\cite{Kalliamvakou:2014}).}

{\footnotesize
\begin{tabular}{r|l}
\rowcolor{gray!30}Check  & Condition  \\\hline
Personal purpose (\# Developers) & $>$ 7 \\
Collaboration (Pull requests) & $>$ 0 \\
Issues & $>$ 10 \\
Releases & $>$ 1 \\
Commits & $>$ 20 \\
Duration & $>$ 1 year 
\end{tabular}}

 \vspace{10mm} 

{\bf    \tbl{data}c: Statistics on selected systems.}

\resizebox{.95\linewidth}{!}{
\begin{tabular}{r@{~}|r@{~}|r@{~}|r@{~}|r@{~}|r@{~}|r@{~}}
\rowcolor{gray!30}    & & Duration & No. of & No. of   & No of. & Reviewed \\
\rowcolor{gray!30} Dataset & Language & (years) & Developers & Commits & Releases & Commits \% \\\hline
LAMMPS & C++ & 5.5 & 84 & 5587 & 7 & 11 \\ 
RMG-PY & Python & 9.5 & 47 & 4472 & 7 & 14 \\ 
PCMSOLVER & C++ & 4.5 & 8 & 1655 & 2 & 17 \\ 
HOOMD & C++ & 3 & 41 & 3904 & 6 & 21\\
ABINIT & Fortran & 2.5 & 23 & 3911 & 9 & 22\\ 
LIBMESH & C & 6.5 & 56 & 7801 & 8 & 25 \\
AMBER & C++ & 4.5 & 11 & 4243 & 4 & 26 \\ 
MDANALYSIS & Python & 4 & 80 & 2733 & 8 & 38 \\ 
XENON & Java & 6 & 11 & 1804 & 7 & 39 \\ 
\end{tabular}
}

\end{center}
\end{minipage} 
\vspace{1.5mm}
~\hrule~
\end{table*}

%% file: commitguru_features.tex
\begin{table*}
\centering
\footnotesize
\caption{14 independent commit-level features, collected by Commit.Guru.\cite{kamei12_jit}.}
\label{tbl:metrics}
\begin{tabular}{l|l|p{4.5cm}|p{9cm}}
\hline
\rowcolor{gray!40}Dimension & Name & Definition & Rationale \\\hline
\multirow{5}{*}{Diffusion} & NS & Number of modified subsystems & Changes modifying many subsystems are more likely to be defect-prone \\\cline{2-4}
& ND & Number of modified directories & Changes touching more directories
are more likely to introduce defect. \\\cline{2-4}
& NF & Number of modified Files & Changes touching more files
are more likely to introduce defect. \\\cline{2-4}
& Entropy & Distribution of modified code across each file & Changes with high entropy are more
likely to introduce technical debt, since a developer will
have to recall and track more scattered changes across
each file. \\\hline
\multirow{3}{*}{Size} & LA & Lines of code added & \multirow{2}{*}{Changing   more lines of code is more likely to introduce
  defects.} \\\cline{2-3}
& LD & Lines of code deleted & \\\cline{2-4}
& LT & Lines of code in a file before the changes & The larger the file/module, the more likely that the change would be defective. \\\hline
\multirow{2}{*}{Purpose}  & FIX & Whether the change is defect  & Changes that fixing the defect are more likely to introduce more defects \\
& & fixing? & than changes for new functionality implementation. \\\hline
\multirow{5}{*}{History} & NDEV & Number of developers that changed the modified files & Changed files touched by more developers before are more likely to introduce defects, since different developers have different design \\
& &  &  thoughts and code styles. \\\cline{2-4}
& AGE & The average time interval from the last to the current change & More recent changes (lower age) contribute more defects than older changes (longer age). \\\cline{2-4} 
& NUC & Number of unique changes to the modified files before & Larger NUC changes are more likely to
introduce defects since a developer will have to recall and track many previous changes. \\\hline
\multirow{5}{*}{Experience} & EXP & Developer experience & The experience of developers has an impact on introducing TD \\\cline{2-4}
& REXP & Recent developer experience & The experience of developers that has often modified the files are less likely to introduce defects (more familiar with the system).  \\\cline{2-4}
& SEXP & Developer experience on a subsystem & Modifications that are made by developer that are familiar with the subsystems are less likely to introduce defects.  \\\cline{1-4}

\end{tabular}
\vspace{15pt}
\end{table*}